\documentclass{IEEEcsmag}

\usepackage[colorlinks,urlcolor=blue,linkcolor=blue,citecolor=blue]{hyperref}
\expandafter\def\expandafter\UrlBreaks\expandafter{\UrlBreaks\do\/\do\*\do\-\do\~\do\'\do\"\do\-}
\usepackage{upmath,color}

\usepackage{doc}
\usepackage{booktabs}
\usepackage{multirow,bigstrut}
\usepackage{tabu}
\usepackage{amsmath}
\usepackage{MnSymbol}
\usepackage{wasysym}
\usepackage{datetime}

\jvol{}
\jnum{}
\paper{}
\jmonth{}
\jname{\hspace{-2.5cm}To appear on IEEE Security \& Privacy, \url{https://doi.org/10.1109/MSEC.2024.3355713}} 
\jtitle{ } 
\pubyear{}

\begin{document}

\sptitle{}

\title{AI Code Generators for Security:\\Friend or Foe?}

\author{Roberto Natella}
\affil{University of Naples Federico II, Naples, Italy}

\author{Pietro Liguori}
\affil{University of Naples Federico II, Naples, Italy}

\author{Cristina Improta}
\affil{University of Naples Federico II, Naples, Italy}

\author{Bojan Cukic}
\affil{University of North Carolina at Charlotte, Charlotte, North Carolina, USA}

\author{Domenico Cotroneo}
\affil{University of Naples Federico II, Naples, Italy}


\markboth{}{}

\begin{abstract}
Recent advances of AI code generators are opening new opportunities in software security research, including misuse by malicious actors. We make the case that cybersecurity professionals need to leverage AI code generators. We review use cases for AI code generators for security, and introduce an evaluation benchmark for these tools.
\vspace*{-0.8cm}
\end{abstract}

\maketitle

\chapteri{L}arge Language Models (LLMs) represent the latest breakthrough in machine learning and are going to have a significant impact on supporting people in various tasks. These models can automatically generate streams of human-like text, as they are trained on huge volumes of text crawled from the web and books, using highly scalable deep-learning architectures. Most notably, these models are also AI code generators, as they can create computer programs using a programming language. Given as input a description of a program in natural language (e.g., plain English), the AI can generate a corresponding program, as a sequence of output tokens.

Computer security is also going to be affected by the advent of these AI code generators. They can represent a new threat, as malicious actors can use them to write new malicious software, bringing in more diversity and agility in their attacks. AI code generators are also easily available to any malicious party, through public services such as GitHub Copilot and Amazon CodeWhisperer, which leverage the same technology behind the well-known ChatGPT and can convert natural language (e.g., in a code comment) into entire methods and functions, from within the development environment.

At the same time, security analysts can (and should) leverage AI code generators. We believe in the need for an open discussion on the uses of this technology for security applications. Since the dawn of the Internet, security analysts have been debating about whether to publicly share information about vulnerabilities and programs to exploit them, since this information can be misused even by inexperienced attackers (e.g., ``newbies'', ``script kiddies''). Attackers will inevitably take any opportunity to use AI code generators; cybersecurity professionals should also strive to benefit from these tools, to better prevent and mitigate intrusions.

The field of generative AI for security is still a young one. Recent studies analyzed this technology in the context of generating malware, malicious content for social engineering, and a few more use cases. However, research on generative AI is limited by the availability of labeled datasets for security use cases, which are needed for fine-tuning an LLM, since these models are only trained in a non-supervised way. Moreover, datasets are needed to support research on new emerging LLMs, by enabling rigorous experimental evaluations.

In this paper, we study the application of AI code generators for creating synthetic attacks. First, we discuss potential benign applications of synthetic attack generation, across many use cases in the context of penetration testing. Then, we present a dataset and an experimental evaluation of three popular LLMs (GitHub Copilot, Amazon CodeWhisperer, and CodeBERT) for generating synthetic attacks. Our novel dataset includes a set of security-oriented programs in Python, which we annotated with descriptions in natural language. Our experiments show that LLMs can be close, but not fully match the accuracy of LLMs at generating general-purpose programs. We found that the best results are obtained with natural language descriptions at a fine-grain (i.e., individual statements rather than whole functions), and by fine-tuning CodeBERT with our dataset. This dataset and experiments can serve as a benchmark for future research.

\section{SECURITY USES (AND MISUSES) OF AI CODE GENERATORS}
\label{sec:uses}
Many professional roles can benefit from AI code generators, including penetration testers, red teams, incident handlers, threat analysts, and more, as all these roles rely on writing custom software to automate complex tasks. Such tasks include the assessment of attack surfaces, the collection and analysis of intelligence, and the emulation of exploitations and adversarial behaviors. 
Moreover, AI code generators can assist newcomers (e.g., students) in writing code for security, since it requires advanced coding skills on how to exploit software vulnerabilities. This barrier is a limiting factor for the growing demand for cybersecurity professionals, which is in need to flatten the learning curve of security-oriented coding and to open to a wider and inexperienced community. 
Thus, ethical hacking can greatly benefit from AI code generators.

Both the defensive and offensive sides invest significant efforts to write programs for automating common tasks, and for scaling for large systems and amounts of data. Scripting programming languages, such as Python, are a typical choice for task automation. These tasks include:

\vspace{3pt}
\noindent
$\blacksquare$ \textbf{Attack Surface Analysis:} The discovery of technical assets that are reachable from outside the target network. Assets include IP addresses, servers, domain names, networks, and IoT objects. These assets are potentially affected by software vulnerabilities and misconfigurations that can be exploited by an attacker. For example, both defenders and attackers can write tools to enumerate subdomains, scan network ports, crawl web pages, and query search engines (e.g., Shodan), to identify vulnerable hosts and services such as code repositories, admin panels, shared files, email and chat servers, which can be prone to data leaks (e.g., source code, authentication tokens) and can allow unauthorized access if not protected. Automated tools accelerate the analysis of multiple types of assets using different sources of data.

\vspace{3pt}
\noindent
$\blacksquare$ \textbf{Open Source Intelligence (OSINT):} The discovery of pieces of information about people in the target organization, such as names, email addresses, phone numbers, and social network accounts, by looking into publicly reachable sources. Again, attacker-written tools can automate web crawling and parsing to extract this information. This information can be leveraged for attacks, such as for social engineering and brute forcing. For example, in brute force attacks, a tool can include personal information to generate tentative usernames and passwords for logging into a system. In social engineering, the attacker can use a tool to craft spear-phishing emails from templates and send them to multiple targets. Similarly, defenders need to collect OSINT to learn about information leaks from their organization, and to perform social engineering attacks for assessment purposes.

\vspace{3pt}
\noindent
$\blacksquare$ \textbf{Vulnerability exploitation:} Attacks rely on automation to trigger vulnerabilities. Once a vulnerability has been discovered, malicious attackers use scripts (``exploits'') to quickly exploit multiple targets (e.g., different organizations, or different hosts in the same organization). Writing exploits are of high interest to security analysts, too. They need exploits to test that their systems are indeed protected from a known attack. Moreover, exploits are often needed to demonstrate the impact and the actual exploitability of a vulnerability (``proof-of-concept'''), to motivate vendors and users to patch them. In the worst case, writing an exploit can show that a vulnerability allows a remote attacker to execute arbitrary code in the target host; in other cases, the attacker may get access to data, cause a denial-of-service (e.g., killing a process, consume resources), or exploit more vulnerabilities. It is challenging for vendors to tell apart vulnerabilities that are indeed exploitable; for example, CVE data are not technically verified, and often misleading.

\vspace{3pt}
\noindent
$\blacksquare$ \textbf{Post-exploitation activities:} Getting a foothold through an exploited vulnerability is only the initial step of an attack (the ``cyber kill chain''). Both attackers and security analysts (``red teams'') leverage automated tools for lateral movement and privilege escalation, by stealing credentials from sniffed traffic or compromised hosts; for persistence, by installing backdoors and remote-control tools to provide them access and maintain it over time (e.g., after reboots); for data theft and exfiltration, by logging keystrokes and screens, and transmitting the stolen data to an external network. Attackers write custom programs for all these activities, to tailor the attack for the specific victim, and for evading anti-virus, network monitoring, and endpoint detection and response (EDR) solutions. For example, malware is often delivered as an encrypted payload, to be launched with a decryption program (an ``unpacker''); attackers apply their own, custom-made (and even simplistic) encryption, to differentiate from other attacks and evade malware detection signatures. Similarly, red teams emulate real attacks using custom-made software, to realistically assess the effectiveness of procedures and solutions for attack detection. Moreover, security analysts can use AI-generated code for automating incident response actions.

These use cases show that offensive security is a software-intensive area, but writing offensive code takes its toll on the time budget. Moreover, it can be a technically difficult activity. For example, in exploit development, ``shellcode'' payloads are typically written in assembly language, to perform low-level operations and to gain full control of the layout of code and data in stack and heap memory, such as to make the shellcode more compact and obfuscated. However, programming in assembly is time-consuming and has low productivity. In testing anti-malware solutions, writing malware requires working with (and abuse of) the complex C++ APIs from the Microsoft Windows OS and related products. Higher-level languages such as Python make it easier to write offensive code but provide less flexibility and can still require a significant amount of time to write code. 

We make the case that security analysts need to leverage AI code generators to get support for defensive tasks. In this case, developers would translate a description of a piece of code in English (an ``intent''), into the corresponding code snippet. For example, security analysts can query the AI for code snippets that they could not recall, that are not yet confident to write themselves, in a similar way to querying a search engine, with the additional benefit that the generated code is tailored for the specific application. Moreover, working with security code, such as in assembly language, can be a barrier for newcomers in security, which is a limiting factor for the growing demand for security analysts able to work with low-level attacks. Thus, AI code generators can flatten the learning curve with natural language processing. Finally, as malicious actors reap the benefits of AI code generators (e.g., to develop more diverse malware in larger quantity), security analysts also need to leverage AI to keep up with the pace.

\section{EXPERIMENTAL EVALUATION}
\label{sec:experiments}

We experimented with AI code generators in the context of several security tasks. For evaluation purposes, we build our own manually curated dataset (\textit{violent-python}\footnote{Available at: \url{https://github.com/dessertlab/violent-python}}), where a sample contains a piece of code from an offensive software (in a programming language), and its corresponding description in natural language (plain English). 

We built the dataset using the popular book ``Violent Python'' by T.J. O'Connor~\cite{o2012violent}, which presents several examples of offensive programs using the Python language. Our dataset covers multiple areas of offensive security, including penetration testing (e.g., an automated exploit for an SMB vulnerability, a port scanner, an SSH botnet); forensic analysis (e.g., geo-locating individuals, recovering deleted items, inspecting the Windows registry, examining metadata in documents and images, and analyzing data from mobile and desktop applications); network traffic analysis (e.g., capturing packets and geo-locating IP addresses, identifying DDoS toolkits, discovering decoy scans, analyzing botnet traffic, foiling intrusion detection systems); OSINT and social engineering (e.g., anonymously browsing the web, working with developer APIs, scraping popular social media sites, creating a spear-phishing email). 

The dataset consists of 1,372 unique samples, as shown in \tablename{}~\ref{tab:table1}. Note that the row total indicates the total number of unique examples (i.e., we did not report replicated pairs of natural language intent/code snippets). 
This dataset is complementary to our previous datasets (\textit{Shellcode\_IA32} and \textit{EVIL}), where we included code snippets from shellcodes in assembly language~\cite{liguori2022can}, and from exploits in mixed Python and assembly language~\cite{liguori2021evil}.

The size of our dataset is in line with other state-of-the-art corpora used to fine-tune ML models. In fact, in state-of-the-art code generation, a model is not trained from scratch, but existing Large Language Models (that were already trained with millions of publicly available lines of code) are fine-tuned in a supervised way, to achieve transfer learning for the specific case (in our case, generating offensive code). Typically, the datasets for fine-tuning are relatively limited, in the order of one thousand samples~\cite{zhou2023lima}.

{
\renewcommand{\arraystretch}{1.5}
\begin{table}[ht]
\caption{The \textit{violent-python} dataset}
\label{tab:table1}
\scriptsize
\begin{tabular}{
>{\centering\arraybackslash}m{2.25cm} |   
>{\centering\arraybackslash}m{1.25cm}
>{\centering\arraybackslash}m{1.25cm}
>{\centering\arraybackslash}m{1.25cm}}
\toprule
 & \textbf{Individual lines} & \textbf{Multi-line blocks} & \textbf{Functions}\\
\midrule
\textit{Penetration Testing} & 490	 & 48 & 21\\
\textit{Forensic Analysis} & 342 & 	47 & 13\\
\textit{Network Traffic Analysis} & 375 & 43 & 20\\
\textit{OSINT and Social Engineering} & 553 & 55 & 25\\
\midrule
\textbf{\textit{Total}} & \textbf{1,129} & \textbf{171} & \textbf{72}\\
\bottomrule
\end{tabular}
\end{table}
}

\begin{table*}[ht]
\caption{Examples of intents in natural language.}
\label{tab:table2}
\scriptsize
\begin{tabular}{
>{\arraybackslash}m{3.5cm} |   
>{\arraybackslash}m{3.5cm} |
>{\arraybackslash}m{4cm} |
>{\arraybackslash}m{3cm}}
\toprule
\textbf{Code} & \textbf{Individual lines description} & \textbf{Multiple lines (block) description} & \textbf{Entire function description}\\
\midrule
\texttt{def connScan(tgtHost, tgtPort)}	& \textit{Define function connScan with parameters tgtHost and tgtPort}		&  & \\ 
& & & \\
\texttt{try:} & \textit{Start try block }	 & \multirow{3}{4cm}{\textit{try to create the socket with parameters AF\_INET and SOCK\_STREAM, connect to tgtHost on tgtPort, send the message ’ViolentPython’, receive the response in result and acquire the lock}} & \\
& & & \\
\texttt{connSkt = socket(AF\_INET, SOCK\_STREAM)}	& \textit{Create the socket with parameters AF\_INET and  SOCK\_STREAM} & & \multirow{3}{3cm}{\textit{Send the message Violent-Python to the host tgtHost on the port tgtPort and receive the response}}\\
& & &\\
\texttt{connSkt.send( ViolentPython \textbackslash r\textbackslash n)}	& \textit{Send the message Violent-Python} & &\\
& & & \\
\texttt{results = connSkt.recv(100)}	& \textit{Receive the response in result} & & \\	
\bottomrule
\end{tabular}
\end{table*}

In our evaluation, we considered several approaches to describe offensive code in natural language, since this is an important factor to determine the usability of AI code generators. One approach is to describe individual lines of code with an English statement, which is typical of other datasets in the field of code generation. This approach can provide the highest accuracy of the generated code since the developer guides the AI code generator with a fine-grained description. However, this approach is also the most verbose and demanding one for the developer. Therefore, we also consider other two approaches where, respectively, we use an English statement to describe multiple lines of code (``blocks'') and entire functions. In the case of blocks and functions, multiple code snippets are joined by the newline character ``\textbackslash n''. 
Overall, the dataset consists of $82\%$ of individual lines, $12\%$ of multi-line blocks, and $6\%$ of entire functions.
For every script in the dataset, we manually described it in the three alternative ways. We based the descriptions on the contents of the chapter around each script, and on comments in the code where available. 
\tablename{}~\ref{tab:table2} shows examples of descriptions of the three alternative granularities.

To evaluate AI code generators for security purposes, we start from  \textbf{CodeBERT}\footnote{\url{https://github.com/microsoft/CodeBERT}}, a pre-trained language model for programming languages. CodeBERT is a model representative of the state-of-the-art, which has achieved high performance in several software engineering tasks, including the generation of offensive code.

It is a multi-programming-lingual model pre-trained on natural language to programming language pairs, across 6 different programming languages. CodeBERT represents the state-of-the-art for several code-related tasks in the software engineering field, such as code search and generation of code and other artifacts, such as comments, documentation, and commit messages. 
According to the best practices for using pre-trained models, we use part of our dataset as training data, to \textit{fine-tune} the model for the specific task of generating offensive code. Moreover, we run the model along with \textit{data processing} operations~\cite{liguori2021evil}, both before translation to prepare the input data, and after translation to improve the quality and the readability of the code in output.
For our experiments, we used a machine with a Debian-based distribution, with 8 vCPU, 16 GB RAM, and two NVIDIA T4 GPUs.

We assessed the model's ability to generate offensive code from different styles of natural language, according to the three different levels of details in the descriptions of the dataset (i.e., line, blocks, and functions). 
We split the dataset into sets for training (the samples for fine-tuning the model), validation (to tune the hyperparameters of the models), and test (for the evaluation), using a random selection with the common $80\%-10\%-10\%$ ratio.

To estimate the correctness of the AI-generate code, the golden standard is represented by a manual code review where a human evaluator checks if the code generated by the models is semantically correct, i.e., it performs exactly what is described in the NL intent. However, manual review is often unfeasible due to the large amount of data to scrutinize, which makes the process time-consuming and prone to errors.

Therefore, the most common practice is to adopt output similarity metrics to assess the similarity of the code generated by the models with a reference ground truth.
Among the large set of available output similarity metrics, we choose the \textit{edit distance} (ED). We based this choice on our previous work \cite{liguori2023evaluates}, where we systematically analyzed several similarity metrics for both Python and assembly code and analyzed the correlation of these metrics with semantic correctness. 
This metric measures the edit distance between two strings, i.e., the minimum number of operations on single characters required to make each code snippet produced by the model equal to a reference code snippet from the dataset, which is used as ground truth for the evaluation. The ED ranges between $0$ and $1$, with higher scores corresponding to smaller distances.

Including metrics that assess if the code is compilable would not bring any useful information since these metrics assess the syntactical correctness rather than the semantic one. In fact, a code can be syntactically correct (i.e., compilable) but still do not perform what is described in the intent (i.e., semantically incorrect). As a matter of fact, metrics such as compilation accuracy have shown to be the less correlated ones to the semantics correctness of security-oriented code, for both the Python and assembly languages~\cite{liguori2023evaluates}.

\begin{figure}
    \centering
    \includegraphics[width=1\columnwidth]{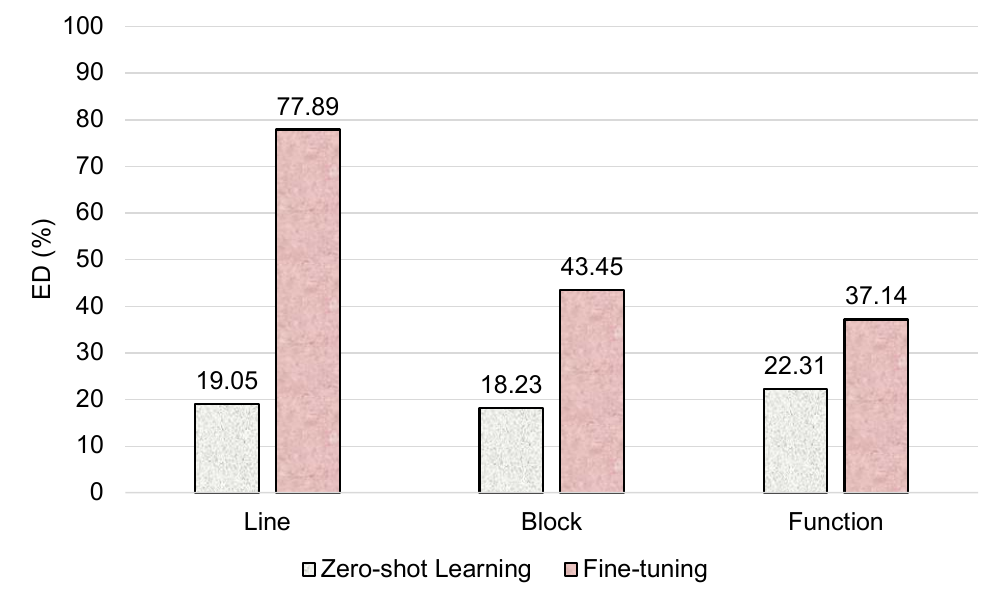}
    \caption{Zero-shot learning versus offensive code fine-tuning.}
    \label{fig:codebert}
\end{figure}

To understand how the model fine-tuning impacts the performance, we compare the results against the performance of the model without any fine-tuning, also known as \textit{zero-shot learning}. \figurename{}~\ref{fig:codebert} shows the results in terms of ED (\%). Unsurprisingly, the results highlight that fine-tuning the model on offensive code always provides performance higher than the one obtained without fine-tuning. The boost in the performance is more evident when the models generate individual lines ($19.05\%$ vs $77.89\%$) and becomes closer when the model generates blocks ($18.23\%$ vs $43.45\%$) and functions ($22.31\%$ vs $37.14\%$).
This happens mainly for two reasons: \textit{i)} the fine-tuned model, as expected, is less accurate at generating complex code (e.g., code blocks and functions) than individual lines; \textit{ii)} the model without fine-tuning (i.e., zero-shot learning) is insensitive to the complexity of code to be generated. Even better, during zero-shot learning, CodeBERT generates functions with higher performance than the one obtained during the generation of blocks and single lines. Most likely, the data used to pre-train CodeBERT contains several examples of complex code such as entire functions rather than simple code snippets.

Then, we compare the fine-tuned CodeBERT against two popular and widely used public AI-code generators: \textbf{GitHub Copilot}\footnote{\url{https://github.com/features/copilot}} and \textbf{Amazon CodeWhisperer}\footnote{\url{https://aws.amazon.com/codewhisperer}}. They are both public services that empower AI code assistants within the development environment, by providing code suggestions from comments in natural language and from existing code. They were trained on billions of lines of code from open-source projects. These solutions are accessible via APIs.

\begin{figure}
    \centering
    \includegraphics[width=1\columnwidth]{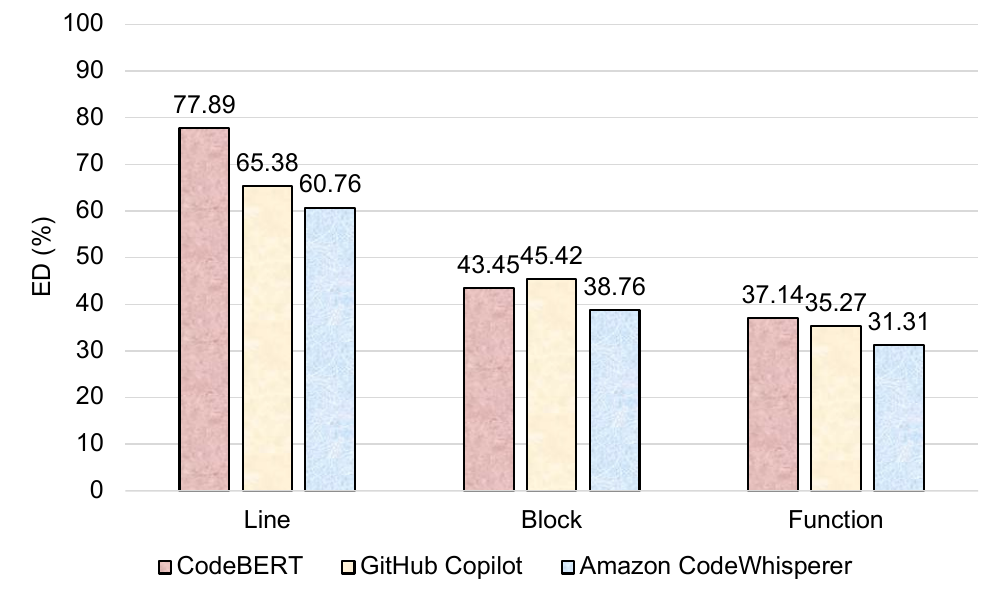}
    \caption{Comparison of AI code generators in the generation of offensive Python code.}
    \label{fig:experiments}
\end{figure}

We compare the performance of the three AI code generators on the same test set. We used the training data only for CodeBERT since it is not possible to further fine-tune public AI code generators. \figurename{}~\ref{fig:experiments} shows the results, in terms of ED (\%), of the AI code generators in the generation of single lines, code blocks, and entire functions of security-oriented Python code. 
First, the figure shows that performance decreases from single lines to code blocks, and from code blocks to entire functions, regardless of the code generator. This is an expected result due to the increasing complexity of the code to be generated. Let's analyze the results thoroughly.
For the simplest task, i.e., the generation of single lines, CodeBERT ($77.89\%$) provides the best performance, followed by Copilot ($65.38\%$) and CodeWhisperer ($60.76\%$).
We attribute this to the fine-tuning of the model since it helps to better perform at generating offensive code. 
For the blocks and functions, where the amount of samples in the dataset is limited, we found that CodeBERT and Copilot have similar performance ($43.45\%$ vs $45.52\%$ for blocks, $37.14\%$ vs $35.27\%$ for functions), hence, the fine-tuning does not boost the performance of CodeBERT when compared to the public AI-code generator. 
CodeWhisperer, instead, shows lower performance than the two competitors ($38.76\%$ and $31.31\%$ for blocks and functions, respectively). The reasons can be attributed to the fact that CodeWhisperer, differently from a general purpose tool such as Copilot, caters first and foremost to development use cases associated with Amazon platforms, such as Amazon Web Services, hence to writing code related to Amazon APIs.

\begin{table*}[ht!]
\caption{Examples of generated code and expected code. Generated code that perfectly matches the reference is \textbf{bold}.}
\label{tab:table3}
\scriptsize
\begin{tabular}{  
>{\arraybackslash}m{1cm} |
>{\arraybackslash}m{2cm} |
>{\arraybackslash}m{2.6cm} |
>{\arraybackslash}m{2.6cm} |
>{\arraybackslash}m{2.6cm} |
>{\arraybackslash}m{2.6cm}}
\toprule
\textbf{Type} & \textbf{Natural Language Intent} & \textbf{Ground-truth Reference} & \textbf{CodeBERT output} & \textbf{GitHub Copilot output} & \textbf{Amazon CodeWhisperer output}\\
\midrule

Line & 
\textit{Assign 0 to seqNum} & 
\texttt{\textbf{seqNum = 0}}  & 
\texttt{\textbf{seqNum = 0}} & 
\texttt{\textbf{seqNum = 0}} & \
\texttt{\textbf{seqNum = 0}}\\ \midrule

Block & 
\textit{If city is not equal to an empty string concatenate the strings city, ',' and country to geoLoc} & 
\texttt{\textbf{if city != '': \textbackslash n  geoLoc = city + ', ' + country}} & 
\texttt{\textbf{if}(cityFile = None): \textbackslash n cities = geoLoc} & 
\texttt{\textbf{if city != '': \textbackslash n    geoLoc = city + ', ' + country} \textbackslash n print(geoLoc)} & 
\texttt{city = "London" \textbackslash n if city: \textbackslash n    geoLoc = city + ", UK" \textbackslash n    print(geoLoc) \textbackslash n else: \textbackslash n print("City is not defined")}\\ \midrule

Function & 
\textit{Connect to the database skypeDB and execute the command "SELECT displayname, skypename, city, country, phone\_mobile, birthday FROM Contacts;"}	 &
\texttt{\textbf{def printContacts (skypeDB): \textbackslash n     conn = sqlite3.connect (skypeDB) \textbackslash n     c = conn.cursor() \textbackslash n     c.execute ("SELECT displayname, skypename, city, country, phone\_mobile, birthday FROM Contacts;" \textbackslash n}}  &
\texttt{\textbf{def} printCookies (cookiesDB): \textbackslash n try: \textbackslash n \textbf{conn = sqlite3.connect} (cookiesDB) \textbf{\textbackslash n c = conn.cursor()} \textbackslash n \textbf{c.execute("SELECT displayname, skypename, city, country, phone\_mobile, birthday FROM Contacts;") \textbackslash n} for row in c: \textbackslash n ...} & 
\texttt{\textbf{def} connect\_skypeDB(): \textbackslash n    try: \textbackslash n        \textbf{conn = sqlite3.connect} ('skypeDB') \textbackslash n        cur \textbf{= conn.cursor() \textbackslash n}        \textbf{cur.execute ("SELECT displayname, skypename, city, country, phone\_mobile, birthday FROM Contacts;") \textbackslash n}        rows = cur.fetchall() \textbackslash n        for row in rows: \textbackslash n ...} & 
\texttt{\textbf{def} connect\_db(): \textbackslash n    conn = sqlite3.connect ('skypeDB') \textbackslash n    return conn}\\  \bottomrule

\end{tabular}
\end{table*}

Finally, to provide context about the results achieved by the AI code generators, we can contrast them with the performance of the state-of-the-art large-language models in the generation of general-purpose (i.e., non-offensive) Python code, which is equal to $86.7\%$~\cite{svyatkovskiy2020intellicode}. This result is comparable with the performance on the generation of individual lines of offensive code. Therefore, current language models provide a good potential for generating offensive code, although they need to be fine-tuned for this task and be guided by fine-grained descriptions from developers. As the scale and complexity of large language models grow, we can expect that their overall performance in security applications will grow over time.

To provide more practical insights into the code generated by the code generators, \tablename{}~\ref{tab:table3} presents a qualitative analysis using cherry-picked examples from our test sets. 
It is important to remark that the code we collected for the dataset is always related to security use cases. Unlike regular code generation that focuses on non-security programs (e.g., CRUD applications, small algorithms on data structures), security-oriented code contains a large number of low-level arithmetic, logic operations, and bit-level slices~\cite{yang2023exploitgen}. Given that security code consists of several lines of code, some of them can appear as ``general purpose'' code if considered out of their context. However, all the examples in our dataset are indeed part of a real security use case.

The first row of the table shows how the models can correctly generate individual line code snippets, by performing a simple operation such as the assignment of a variable. 
The second row shows the ability of Copilot in the generation of a whole multi-line block, by generation a correct sequence of an if-then statement. CodeBERT and Amazon CodeWhisperer, instead, both fail to generate the correct output. Indeed, the former generates something close to the expected output, yet incomplete. The latter, on the other hand, produces a verbose snippet that is syntactically correct but that diverges from the natural language description. 
Finally, the third row shows how the models deal with the generation of whole, complex functions from a single natural language description. CodeBERT and Copilot prove all their potential by generating several lines of code snippets, hence, they prove to be a valid solution as code assistants for more complex tasks. 
The code generated by CodeWhisperer, again, generates a syntactically correct function that, however, contains only a subset of the operation required to accomplish what is required by the natural language prompt.

\section{RELATED WORK}
\label{sec:related}
Given their recent advances, AI-based solutions have become an attractive solution for different tasks in the field of software security. \tablename{}~\ref{tab:related} presents the related work.

\begin{table}[ht]
\caption{Related Work.}
\label{tab:related}
\scriptsize
\begin{tabular}{
>{\arraybackslash}m{0.5cm} |   
>{\arraybackslash}m{1.25cm} |
>{\arraybackslash}m{4.75cm} 
}
\toprule
\textbf{Year} & \textbf{Author} & \textbf{Contribution}\\
\midrule
2022 & Liguori \textit{et al.}~\cite{liguori2022can} & Use of NMT models to generate software exploits in assembly language from NL. \\
\midrule
2022 & Kim \textit{et al.}~\cite{kim2022eco} & Security surveillance toward AI-enabled digital twin service, which provides eco-friendly security by the active participation of low-resource devices.\\
\midrule
2022 & Yang \textit{et al.}~\cite{yang2022dualsc} & 
Use of a shallow transformer model that performs code generation and summarization to generate software exploits.\\
\midrule
2023 & Yang \textit{et al.}~\cite{yang2023exploitgen} & Generation of software exploits using a rule-based template parser to generate augmented NL descriptions and a semantic attention layer to extract and calculate each layer’s representational information.\\
\midrule
2023 & Xiao \textit{et al.}~\cite{xiao2023specializing} & Use of neural network-based API completion techniques to capture program dependencies.\\
\midrule
2023 & McIntosh \textit{et al.}~\cite{mcintosh2023harnessing}& Use of a state-of-the-art large language model in generating cybersecurity policies to deter and mitigate ransomware attacks that perform data exfiltration.\\
\midrule
2023 & Pa \textit{et al.}~\cite{pa2023attacker}& Development of malware programs and attack tools using public AI-generative models.\\
\midrule
2023 & Gupta \textit{et al.}~\cite{gupta2023chatgpt} &  Use of public AI code generator to create social engineering attacks, phishing attacks, automated hacking, attack payload generation, and malware. \\
\midrule
2023 & Botacin \textit{et al.}~\cite{botacin2023gpthreats} & Use of public AI code generator to generate malware.\\
\midrule
2023 & Grigoriadou \textit{et al.}~\cite{grigoriadou2023hunting} & Detection and mitigation of IoT cyberattacks by using an AI-powered intrusion detection and prevention system.\\
\bottomrule
\end{tabular}
\end{table}

Our work uses AI code generators for the generation of offensive code.
Differently from previous work on generative AI, we adopt AI-based code generators to support several types of synthetic attacks in the context of penetration testing. Indeed, due to the lack of corpora containing security-oriented code to train AI-based solutions, previous work focused on other, more specific use cases, such as the generation of exploits with low-level languages,  malware generation, and malicious contents for social engineering. Therefore, our work aims to expand the scope of generative AI for security, by introducing a novel dataset and experimental baselines for research in this area.

\section{CONCLUSION}
\label{sec:conclusion}
In this experience with LLMs, and in building a security-oriented evaluation benchmark, we learned about potential use cases for offensive security. These use cases encompass attack surface analysis, OSINT, vulnerability exploitation, and post-exploitation. We believe that cybersecurity professionals must embrace AI code generators to prevent attacks more efficiently.

Overall, the results of our experiments on current AI code generators emphasize the importance of a careful choice of which one to use. 
In particular, the experiments showed the importance of fine-tuning the models for security-oriented applications. In fact, when trained with samples of security-oriented code, CodeBERT can outperform popular public AI code generators, and achieve performance close to non-security-oriented applications. 
Unfortunately, the availability of data for security applications is very limited, and the creation of corpora from scratch is a difficult and time-consuming task, as it requires a significant manual effort supported by high expertise and technical skills in the field. 

When security-oriented code to fine-tune the models is not available, the usage of public AI code generators is a potential solution, although with a performance loss. The choice of the public generator strongly depends on the application, but the experiments we performed suggest that a general-purpose usage tool, such as GitHub Copilot, that was trained with more diverse programming languages and projects, can better deal with the generation of offensive code than generators tailored for specific APIs and architectures, such as Amazon CodeWhisperer.

\section{ACKNOWLEDGMENTS}
This work has been partially supported by MUR PRIN 2022, project FLEGREA, CUP E53D23007950001 (\url{https://flegrea.github.io/}).

\def\refname{REFERENCES}

\bibliographystyle{IEEEtran}
\bibliography{main}

\end{document}